# A SMALL FULLERENE ($C_{24}$) MAY BE THE CARRIER OF THE 11.2 MICRON UNIDENTIFIED INFRARED BAND

L. S. Bernstein[1], R. M. Shroll[1], D. K. Lynch[2], & F. O. Clark[3]

[1]Spectral Sciences, Inc., 4 Fourth Ave., Burlington, MA 01803, USA; larry@spectral.com, rshroll@spectral.com
[2]Thule Scientific, P. O. Box 953, Topanga CA 90290, USA; dave@caltech.edu
[3]Wopeco Research 125 South Great Road, Lincoln, MA, 01773, USA; frank.clark@gmail.com

## ABSTRACT

We analyze the 11.2 μm unidentified infrared band (UIR) spectrum from NGC 7027 and identify a small fullerene ($C_{24}$) as a plausible carrier. The blurring effects of lifetime and vibrational anharmonicity broadening obscure the narrower, intrinsic spectral profiles of the UIR band carriers. We use a spectral deconvolution algorithm to remove the blurring, in order to retrieve the intrinsic profile of the UIR band. The shape of the intrinsic profile, a sharp blue peak and an extended red tail, suggests that the UIR band originates from a molecular vibration-rotation band with a blue band head. The fractional area of the band-head feature indicates a spheroidal molecule, implying a non-polar molecule and precluding rotational emission. Its rotational temperature should be well approximated by that measured for non-polar molecular hydrogen, ~825 K for NGC 7027. Using this temperature, and the inferred spherical symmetry, we perform a spectral fit to the intrinsic profile that results in a rotational constant implying $C_{24}$ as the carrier. We show that the spectroscopic parameters derived for NGC 7027 are consistent with the 11.2 μm UIR bands observed for other objects. We present density functional theory (DFT) calculations for the frequencies and infrared intensities of $C_{24}$. The DFT results are used to predict a spectral energy distribution (SED) originating from absorption of a 5 eV photon, and characterized by an effective vibrational temperature of 930 K. The $C_{24}$ SED is consistent with the entire UIR spectrum and is the dominant contributor to the 11.2 and 12.7 μm bands.

## 1. INTRODUCTION

The unidentified infrared bands (UIR bands) are a family of six major interstellar emission features in the 3-14 μm range (see Figure 1a), appearing together in many different galactic and extragalactic environments (Tielens 2013; Kwok 2011). These features arise from carriers that are linked to many essential processes in our universe, including the form, function, and fate of carbon-rich material, the infrared radiation budget, and the origin and evolution of complex molecules. Since their discovery in the 1970s (Gillett et al. 1973; Russell et al. 1975) many molecular and particulate carriers have been proposed. It is widely believed that they originate from gas phase carbonaceous molecules; however, the identities of these molecules is not firmly established (Tielens 2008; Allamandola et al. 1989; Puget & Leget 1989; Duley & Williams 1981; Peeters et al. 2002; Kwok & Zhang 2011; Bernstein & Lynch 2009).

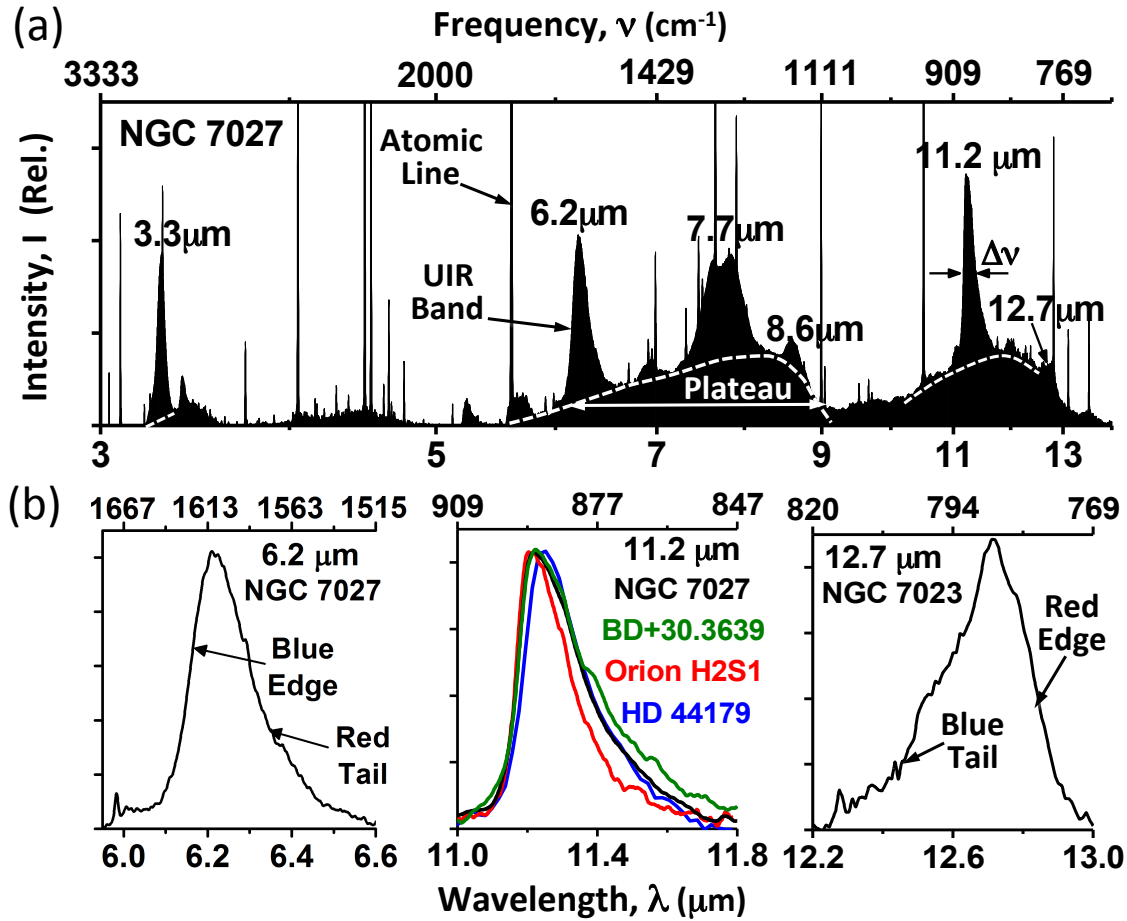

**Figure 1.** Example UIR band spectra for a variety of nebulae. The planetary nebulae NGC 7027, BD +30.3639, Orion H2S1, and HD 44179 spectra were obtained using the ISO-SWS sensor (Candian & Sarre 2015: Sadjadi et al. 2015) and the reflection nebula NGC 7023 spectrum was obtained using the *Spitzer*/IRS sensor (Shannon et al. 2016). **a**. The observed infrared spectrum for NGC 7027. This spectrum rests on a much broader dust continuum (not shown), which has been removed in order to highlight the classic UIR features. The UIR spectrum consists of broad plateaus and narrower UIR bands, for which the central wavelengths are denoted. For several of the features, a dashed white line indicates the estimated boundary between the band and underlying plateau feature. The numerous unresolved line features are due to well-known atomic emission lines. **b**. Several of the UIR bands are shown on an expanded scale after subtracting the underlying plateau. The features show similar shapes, consisting of a steep red or blue edge and a corresponding tail. However, their widths, $\Delta \nu$ (Full Widths Half Maximum - FWHM), are dissimilar, 0.16 μm (43 cm$^{-1}$), 0.20 μm (16 cm$^{-1}$), and 0.26 μm (16 cm$^{-1}$) for the 6.2, 11.2, and 12.7 μm bands, respectively. The plateau-subtracted 12.7 μm feature for NGC 7027 is very noisy; instead, we display a cleaner 12.7 μm spectrum from NGC 7023. The 11.2 μm bands for NGC 7027 and several diverse astronomical objects are compared A smaller, uncorrelated feature, centered at 11.0 μm, was removed, in order to highlight the similarity of the steep blue edges.

The shapes of the UIR bands, typified by the planetary nebula NGC 7027 in Figure 1a, are consistent across most astronomical sources. Amongst those that do differ, much less variation is seen for the 11.2 μm band (Fig. 1b) (van Diedenhoven et al. 2004). The similarity of the 11.2 μm spectra is noteworthy, particularly the near constancy of the steep blue edge. This suggests a single molecule origin, because a molecular band edge would be expected to be nearly invariant for diverse environments. The location of a band edge, usually referred to as a band head, is an invariant spectroscopic property of a molecular vibration-rotation band (Herzberg 1989). It has some environmental dependence in that its appearance requires a sufficiently high rotational temperature to populate its associated rotational energy levels. The focus of this paper is on the exploration of the idea that the 11.2 μm UIR band originates from a vibration-rotation profile of a single molecule.

Most of the UIR bands share a common morphology: a relatively steep blue or red edge with a corresponding, slowly declining red or blue tail. This suggests a single or dominant carrier for each band displaying this shape. The orientation, blue or red, of the edge and tail depends on the sign of the band-specific vibration-rotation constant. Figure 1 highlights the similarity of the 6.2, 11.2, and 12.7 μm band profiles. This similarity extends to other bands, including the strong band near 3.3 μm and the weaker bands at 5.25 and 5.65 μm. It likely extends to more bands, such as those comprising the strong and broad 7.7 μm complex, but the effects of band overlap obscure the similarity of the component bands. It is also evident in non-UIR spectra such as the unidentified 21 μm feature (Volk & Kwok 1999; Hrivnak et al. 200; Zhang et al. 2009). We note that this feature and the NGC 7027 6.2 μm UIR band have nearly identical shapes when overlaid in frequency (i.e., cm$^{-1}$) space. This suggests they may arise from the same carrier. Only one previous study, by Roche et al. 1996 considered the implications of the similarity between the 3.3, 5.25, ad 11.2 μm bands. They suggested that the carriers were highly symmetric molecules, and they noted that some PAHs and fullerenes could satisfy this requirement.

The most widely accepted models attribute the UIR spectrum to a superposition of many, broadened, large polycyclic aromatic hydrocarbons (PAHs), typically ~30, with an average size of ~50 carbon atoms (Tielens 2008). Recent work on the "grandPAH" hypothesis (Andrews et al. 2015) suggests that most UIR spectra may arise from a comparable number of robust PAHs that are abundant in diverse environments. The PAH approach yields good spectral fits to all the classes (i.e., shape variations) of the UIR spectrum, but it has proven challenging to verify. Zhang et al. 2015 argue that it a non-falsifiable model because its large and diverse PAH database can also fit a wide variety of non-PAH spectra. It is inherently difficult to detect a specific PAH due to the obscuring effects of spectral broadening and overlapping spectral bands. The PAH model also suffers from non-uniqueness, as highlighted by two studies that fit a variety of 11.2 μm UIR spectra using different molecular databases (Candian & Sarre 2015; Sadjadi et al. 2015). In contrast, we propose that each UIR band is dominated by the broadened band profiles of a few or even just a single molecular species, with each dominant molecule consisting of less than thirty carbon atoms. Unlike the PAH model, our model provides a physical explanation for the prevalence of the distinctive edge-tail shape displayed by many of the UIR bands, i.e., vibration-rotation band structure.

Only a few previous studies have considered that the UIR band profiles arise from rotational structure. Bernstein & Lynch 2009 explored the possibility that the UIR bands arise from small carbonaceous molecules with $N_C$<10. The results were inconclusive, as only semi-quantitative agreement with band positions and shapes were demonstrated. Le Coupanec et al. 1998 ruled out rotational structure from large, $N_C$>100, PAHs with high, suprathermal rotational temperatures. Pech et al. 2002 argued that rotational structure is not an important contribution to the UIR profiles for $N_C$≥50.

The blurring effects of lifetime and vibrational anharmonicity broadening obscure the narrower, intrinsic spectral profiles of the UIR band carriers.  We introduce the use a spectral deconvolution algorithm to remove the blurring, in order to retrieve the intrinsic profile of a UIR band.  The deconvolution algorithm is applied to the NGC 7027 11.2 μm UIR band.  Its intrinsic profile exhibits finer spectral details than are evident in the observed UIR band profile.  The retrieved intrinsic profile provides tight constraints on the spectroscopic properties of a candidate carrier, such as its shape and size.  The constraints are unbiased; they arise from a single, general assumption – i.e., that the observed carrier spectra are blurred by several forms of broadening, common to all highly excited polyatomic molecules.

The paper is organized as follows.  We first apply a spectral deconvolution algorithm to the NGC 7027 11.2 μm UIR band and show that the shape of the intrinsic profile indicates that the carrier has a spheroidal geometry.  Inclusion of a size constraint, $N_C<30$, based on the width of the band, leads to $C_{24}$ as the only plausible carrier.   Next, a spectral model is used to fit the intrinsic profile and retrieve spectroscopic parameters that corroborate the fullerene $C_{24}$ as the likely carrier.  Then, we show that the $C_{24}$ spectroscopic parameters are consistent with the 11.2 μm spectra for BD +30.3639, HD 44179, Orion H2S1, and NGC 7023, when allowing for variations in the rotational temperature and broadening.  The discussion section features comparison of a predicted spectral energy distribution  (SED) for $C_{24}$ based on density functional theory (DFT) calculations of its vibrational frequencies and infrared intensities, to the NGC 7027 UIR spectrum.  It also includes spectral modeling of the NGC 7023 12.7 μm band and a discussion of remaining challenges to the $C_{24}$ carrier hypothesis. Recommendations are presented for further modeling, laboratory, and observational studies to evaluate the plausibility of the $C_{24}$ carrier hypothesis.  We conclude with a summary of key findings.

## 2. ANALYZING THE 11.2 μm UIR BAND

### *2.1 Spectral Deconvolution of the NGC 7027 11.2 μm Band*

In order to apply a spectral deconvolution algorithm one needs to specify a mathematically well-defined broadening function.  We assume Lorentz broadening is the major broadening component, which is consistent with the prevalent view that the UIR bands arise from highly vibrationally excited polyatomic molecules (Boulanger et al. 1999; Tielens 2008).  These molecules undergo rapid intramolecular energy transfer, due to a high density of vibrational states.  A short lifetime equates to uncertainty in the transition energy, often referred to as Heisenberg broadening, which is described by the Lorentz line shape.  Lorentz broadening is not tied to a specific type of molecule; it is a fundamental behavior of all highly excited polyatomic molecules.

We use Gaussian broadening to approximate the combined effects of the other significant broadening sources.  In descending order of importance, these include vibrational anharmonicity, sensor spectral resolution, and Doppler shifts.  Vibrational anharmonicity arises from coupling between vibrational modes, resulting in shifting of the band origin, usually to the red, and to broadening of the band.  The amounts of shifting and broadening are proportional to the total vibrational energy in a molecule.

We recognize the approximate nature of using a single Lorentz width to represent the lifetime broadening.  As a molecule emits energy, its density of states decreases, which is accompanied by a decrease in its lifetime broadening.  Thus, a more rigorous approach would be to represent the lifetime broadening as an emission intensity weighted sum of Lorentzians with different widths. Since the sum of Lorentz functions is not a Lorentz function, our assumption of a single, effective Lorentzian should be considered as a convenient, but approximate, deconvolution kernel. As shown below, the shape of the retrieved intrinsic profile is not overly sensitive to the values of the assumed Lorentz and Gaussian widths or to the relative contributions of the two forms of broadening. This suggests that the use of approximate deconvolution kernels is sufficient to reveal the key characteristics of the intrinsic profile.

Deblurring an observed UIR band spectral profile is done using deconvolution. We assume a UIR band spectrum, $I_{UIR}$, is related to its intrinsic profile, $I_0$, through the convolution integral,

$$I_{UIR}(\nu) = \int_{-\infty}^{\infty} I_0(\nu'; \gamma_G) L(\nu - \nu'; \gamma_L) d\nu', \tag{1}$$

where $\gamma_G$ is an effective Gaussian width, $\gamma_L$ is Lorentz width, and $L$ is the Lorentz broadening function,

$$L = \frac{1}{\pi} \frac{\gamma_L}{(\nu' - \nu)^2 + \gamma_L^2} . \tag{2}$$

We approximate the convolution integral as a discrete sum with an evenly spaced frequency grid, and use a least squares optimization approach (Byrd et al. 1995; Zhu et al. 1997) to retrieve the intrinsic profile, $I_0(\nu; \gamma_G)$. We can apply a subsequent deconvolution, using a Gaussian function to remove the additional (i.e., non-Lorentzian) broadening effects and retrieve an estimate of the "fully" unbroadened intrinsic molecular vibration-rotation profile. Mathematically, the order in which the deconvolutions are applied does not affect the final result.

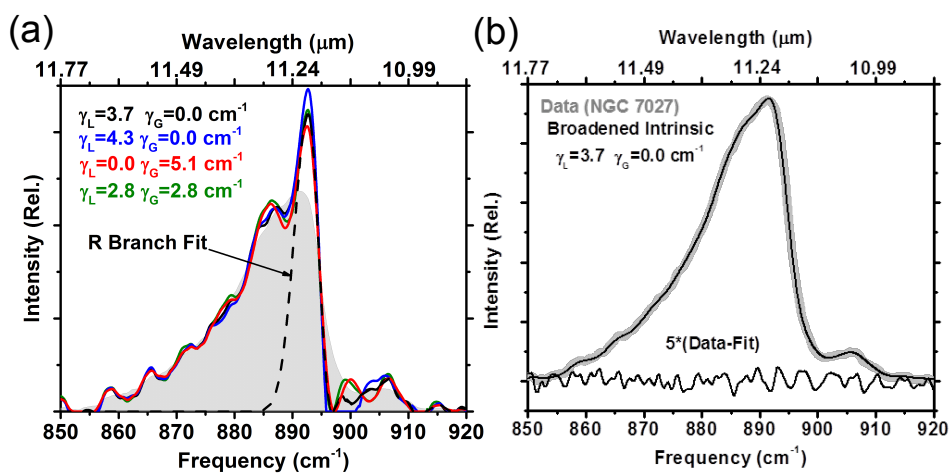

**Figure 2.** Application of the deconvolution algorithm to the NGC 7027 11.2 μm UIR band in frequency (cm$^{-1}$) units. a. Retrieved intrinsic profiles for different choices of the Lorentz, $\gamma_L$, and Gaussian, $\gamma_G$, broadening widths (FWHM). The NGC 7027 data (filled in light grey region) is shown to aid in visualizing the sharpening effects of the deconvolution. A fit to the narrow R-branch feature is also shown (dashed line) which is relevant to constraining the carrier shape (see Sec. 2.3) **b.** Comparison of the broadened intrinsic profile to the original data with the error shown on a 5x expanded scale at the bottom of the plot. A valid intrinsic profile will closely reproduce the original data when it is broadened by the same widths used in its associated deconvolution. As discussed in the text, if unphysical choices are made for the widths, then the errors will be much larger than that shown in 2b.

Figure 2 displays the intrinsic profiles for different choices of the Lorentz and Gaussian broadening widths. The selection of $\gamma_L$=3.7 $\gamma_G$=0.0 cm$^{-1}$ (Figure 2a black curve) corresponds to Lorentz only broadening for which the steep edge first touches the baseline (estimated from the plot at 896 cm$^{-1}$). The selection of $\gamma_L$=4.3 $\gamma_G$=0.0 cm$^{-1}$ (Figure 2a blue curve) corresponds to the maximum Lorentz beyond which the broadened intrinsic profile no longer provides a close fit to the blue tail and peak region of the data (see Figure 2b error curve). The selection of $\gamma_L$=0.0 $\gamma_G$=5.1 cm$^{-1}$ (Figure 2a red curve) corresponds to Gaussian only broadening for which the steep edge first touches the baseline. The selection of $\gamma_L$=2.8 $\gamma_G$=2.8 cm$^{-1}$ (Figure 2a green curve) corresponds to equal Lorentz and Gaussian broadening widths for which the steep edge first touches the baseline. All the intrinsic profiles give comparable errors when broadened and compared to the data.

The key characteristics of the intrinsic profiles are insensitive to significant variations in the selection of the broadening parameters. These include a steep blue edged, and narrow, band head-like feature, followed by a shallow minimum that transitions into a slowly declining red tail. The heights and widths of the narrow features for the different broadening assumptions vary in a manner that preserves the area of the feature.

The intrinsic profiles exhibit several narrow features below ~880 cm$^{-1}$. They correspond to small features in the data that become magnified due to the deconvolution process. The shapes and widths of these features are not similar to the overall profile shape; thus, they are not associated with the underlying molecular vibration-rotation profile. These features are ignored in the spectral fitting of the intrinsic profile (see Sec. 2.4) by constraining the spectral fit to pass below them.

## 2.2 Constraining the Carrier Size

The width of a UIR band can be used to obtain an estimate of its carrier's size. Pech et al. 2002 have derived the following approximate relationship, for a compact PAH, between the width of a vibration-rotation band, $\Delta v$ (cm$^{-1}$), and $BT_{rot}$, the product of the rotational temperature, $T_{rot}$, and the rotational constant, $B$,

$$\Delta v \approx 4\sqrt{BT_{rot}}. \qquad (3)$$

In order to determine $B$ (cm$^{-1}$), and, therefore, obtain an estimate of the size of the molecule, we need to obtain an independent estimate of $T_{rot}$ (K). We argue that an upper limit estimate to $T_{rot}$ is the gas kinetic temperature. We rule out higher temperatures attributable to suprathermal excitation (Le Coupanec et al. 1998), which arises from conversion of absorbed UV photons to high levels of rotational excitation. This follows because the rotational envelopes of the UIR bands are insensitive to orders of magnitude variation in the intensity of the local UV radiation field (Mattila et al. 1996; van Diedenhoven et al. 2004). An upper limit to $T_{rot}$ sets a lower limit to $B$; this corresponds to an upper limit estimate of the moment of inertia, which is a measure of molecular size – i.e., $N_C$. It is well established that in environments containing the UIR bands a good measure of the gas temperature is provided by the rotational temperature of molecular hydrogen, $T_{H2}$, determined from its lowest few rotational levels (Habart et al. 2005; Bernard-Salas & Tielens 2005). Evaluation of eq.(3) for the NGC 7027 11.2 μm band, using $\Delta v$=16 cm$^{-1}$ (see Figure 1b) and $T_{H2}$=825 K (Tielens 2005), yields $B$=0.019 cm$^{-1}$. This value of $B$ corresponds to a PAH with $N_C$=18. The size estimate depends on the assumed molecular geometry. We have applied the approach of Pech et al. 2002 to linear and spherical molecules and find $N_C=6$ and 27, respectively. Based on these estimates for $N_C$ we restrict our search for the carrier of the NGC 7027 11.2 µm band to $N_C<30$.

*2.3 Constraining the Carrier Shape and Identifying the Carrier*

The shape of the intrinsic profile, specifically the fractional area under the R-branch feature, provides a constraint on the carrier geometry. This geometry constraint, when combined with the size constraint, leads to a unique carrier identification. In spectroscopic terms, the blue band head feature corresponds to an R branch. It is well established that for symmetric top parallel band transitions (A→A transition symmetry) the fractional area of the R-branch feature is a direct measure of molecular geometry via the ratio of the rotational constants, *B/C* (see dashed line in Figure 3) (Herzberg 1989). Because the corresponding relationship for perpendicular bands (A→E) (see solid line in Figure 3) was not readily available, we empirically determined it using the PGOPHER spectral simulation code (Western 2012). For perpendicular bands of oblate symmetric tops the fractional R-branch area is constant (i.e., 1/3), independent of the *B/C* ratio. Thus, one cannot distinguish between a spheroidal fullerene and an oblate symmetric top PAH molecule. In this case, additional information, such as whether or not a candidate carrier has an A→E 11.2 µm band, would be required to resolve the shape ambiguity.

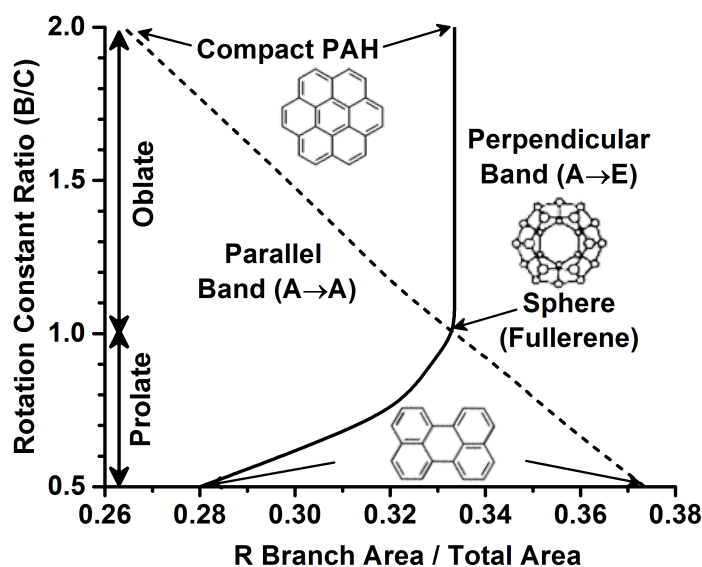

**Figure 3.** Relationship of the ratio of the *B* to *C* rotational constants for a symmetric top molecule to the fractional area of the R-branch for perpendicular (solid line) and parallel (dashed line) transition symmetries. The *B* and *C* constants correspond to rotations about an axis perpendicular or parallel to the symmetry axis, respectively. The perpendicular and parallel transition symmetries correspond to transition moments perpendicular or parallel to the symmetry axis, respectively. Also shown are examples of oblate (top, Coronene, $C_{24}H_{12}$, B/C=2.0), spheroidal (middle, $C_{24}$, B/C=1.0), and prolate (bottom, Perylene, $C_{20}H_{12}$, B/C=0.5) molecules in the size range of $N_C<30$.

The fractional area of the R-branch feature was estimated using the spectral fit presented in Figure 2a. We used a Gaussian function peaking at 892.7 cm$^{-1}$, with a width of 2.75 cm$^{-1}$ (FWHM), to model the red half of the feature. The intrinsic profile was used for the blue half of the feature. The estimated fractional area of the R-branch feature is 0.33±0.02. As discussed earlier, the small feature below 880 cm$^{-1}$ are assumed to be uncorrelated to the primary carrier of the 11.2 μm band; hence, we reduced the total area by 4% to remove their contribution. The uncertainty estimate is based on considering several different functional forms for the red half of the feature, including a linear decline and a Gaussian multiplied by a power law.

The range of values for the fractional area of the R-branch feature, 0.31-0.35, rules out any prolate PAH conforming to our size constraint, $N_C$<30, as a potential carrier. These prolate PAHs have $B/C$<0.5 (Malloci et al. 2007), and, according to Figure 3, they have fractional areas either below 0.28 or above 0.37. These fractional areas fall outside the permissible range of 0.31-0.35. The only potential compact PAH carrier with $N_C$<30 and with a plausible 11.2 μm feature is Coronene ($C_{24}H_{12}$ which has B/C = 2.0, – see Figure 3 for the structure) (Boersma et al. 2014). However, it has A→A symmetry and is therefore not a viable carrier - i.e., its 11.2 μm feature has a fractional R-branch area of 0.26 (see dashed line in Figure 3). This limits our viable carrier options to one of the smaller, spheroidal fullerenes (i.e., $C_{20}$, $C_{24}$, $C_{26}$, or $C_{28}$).

Recent DFT calculations by Adjizian et al. 2016 for the vibrational frequencies and IR intensities of the small fullerenes show that only one, $C_{24}$, has a strong feature close to the 11.2 μm UIR band and additional IR features that fall plausibly close to the other UIR bands. In a later section, we present the results from our DFT calculations that are in reasonable agreement with Adjizian et al. 2016. Our calculations show that the predicted 11.2 μm band has the required A→E transition symmetry. They also enabled a prediction of the $C_{24}$ SED that is shown to be consistent with the NGC 7027 UIR spectrum. In the following section, we show that spectroscopic modeling of the intrinsic profile also supports the identification of $C_{24}$ as the carrier.

## 2.4 Spectral Modeling of the Intrinsic Profile

The spectral fit parameters included the rotational constant, $B$, the difference between the upper and lower vibrational state rotational constants, $\Delta B$, rotational temperature, $T_{rot}$, band origin, $\nu_0$, and the Lorentz and Gaussian widths, $\gamma_L$ and $\gamma_G$. As noted earlier, for NGC 7027 we constrain $T_{rot}$ to be equal to $T_{H2}$=825 K. We first fit the intrinsic profile, which provides good initial values for $B$, $\Delta B$, $\nu_0$, and $\gamma_G$ (the effect of $\gamma_L$ was removed by the deconvolution). We then fit the broader observed profile, allowing for small perturbations of the initial fit values, and also including $\gamma_L$ in the fitting process. The width used in the deconvolution provides a good initial estimate of $\gamma_L$. The fitting was performed visually; each parameter affects the modeled spectrum in a unique manner, easily seen when overlaid on the data. The fitting method also made it easy to disregard features that are not associated with the dominant carrier of the UIR feature, such as the one at 905 cm$^{-1}$ and the three smaller features below 880 cm$^{-1}$. Because these features contribute intensity on top of the dominant feature, we negate their effect by constraining potential fits to pass below them. The estimated fit uncertainties are: $\delta\nu_0$=±0.1 cm$^{-1}$, $\delta BT_{rot}/BT_{rot}$=±5 %, $\delta\Delta B/\Delta B$=±10 %, $\delta\gamma_G$ =±0.2 cm$^{-1}$, and $\delta\gamma_L$ =±0.2 cm$^{-1}$. The spectral fits to the intrinsic profile and the observed data for NGC 7027 are shown in Figure 4, and the fit parameters can be found in Table 1.

Our earlier, tentative identification of $C_{24}$ is corroborated here through the values of the spectral fit parameters, in particular the rotational constant, $B$=0.0155 cm$^{-1}$ and its vibrational energy dependence, $\Delta B/B$=-0.363 %. In order to compare to theoretical rotational constants for $C_{24}$, based on DFT calculations discussed later, we need to correct the retrieved rotational constant to its ground vibrational state value. We estimated this value in the following manner. We assumed an excitation energy of 5 eV, which is typical for aromatic molecules, like PAHs and fullerenes. The effective number of vibrational excited states produced by absorption of a 5 eV photon is ~40, assuming the median energy of a vibrational mode ~1000 cm$^{-1}$. If we associate the average value of the rotational constant, which approximately corresponds to the retrieved value, with the median level of vibrational excitation (i.e., 50% of the energy left in the molecule), then the rotational constant correction factor is ~+20*0.363 %=+7.3 %. This means that the ground state rotational constant is ~1.073*0.0155=0.0166 cm$^{-1}$. The DFT calculations show that $C_{24}$ is a slightly oblate symmetric top with $B$=0.0181 and $C$=0.0160 cm$^{-1}$, corresponding to an average value of 0.0171 cm$^{-1}$ (i.e., $(2B+C)/3$). Therefore, the rotational constant retrieved from the spectral fit is consistent with the ground state rotational constant for $C_{24}$.

As noted above, $C_{24}$ is not a perfect sphere. However, the differences in modeling the spectra with a sphere versus a slightly oblate top are subtle and indistinguishable in the UIR data due to the large amount of broadening.

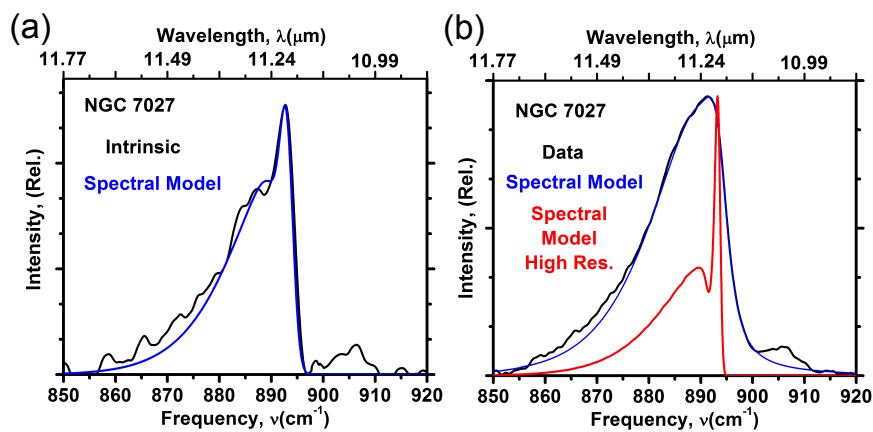

**Figure 4.** Spectral modeling of the NGC 7027 11.2 μm profile. **a.** Intrinsic spectrum after deconvolving Lorentzian broadening (black) and the fitted spectral model based on a spherical molecule (blue; see Table 1 for spectroscopic fit parameters). **b.** Comparison of the spectral fit to the original observed data (blue; a refinement of the model after applying Lorentz broadening to the spectral fit in 4a). Also shown is a high spectral resolution model profile with most of the broadening removed ($\gamma_L$=0.0 and $\gamma_G$=1.0 cm$^{-1}$, shown at a reduced scale to keep the band-head peak within the plot). The model calculations were performed with the Program for Simulating Rotational Structure (PGOPHER) computer code in which the $C$ top axis is taken to coincide with the symmetry axis.

*2.5 Spectral Modeling of the 11.2μm Bands for NGC 7023, Orion H2S1, HD 44179, and BD +30.3639*

We used the spectroscopic parameters retrieved for NGC 7027 as an initial guess for fitting the11.2 μm bands for the other objects, NGC 7023, Orion H2S1, HD 44179, and BD +30.3639. We selected these objects because several groups have recently performed detailed studies of their 11.2 μm bands. We used the same fitting approach as discussed above for NGC 7027; we varied the NGC 7027 spectroscopic fit parameters to fit the observed data. There was no need to perform a deconvolution for these objects, since the NGC 7027 spectroscopic parameters provided a good initial guess. We allowed all the parameters, except for $B$, to vary. We assumed that $B$ was well determined from the NGC 7027 analysis, and that variation in $BT_{rot}$, was dominated by variations in $T_{rot}$. The fits are displayed in Figure 5 and the fit parameters can be found in Table 1. As discussed later, the variability in the molecular spectral fit parameters can be tied to variations in the local physical conditions, such as gas temperature and the radiation field.

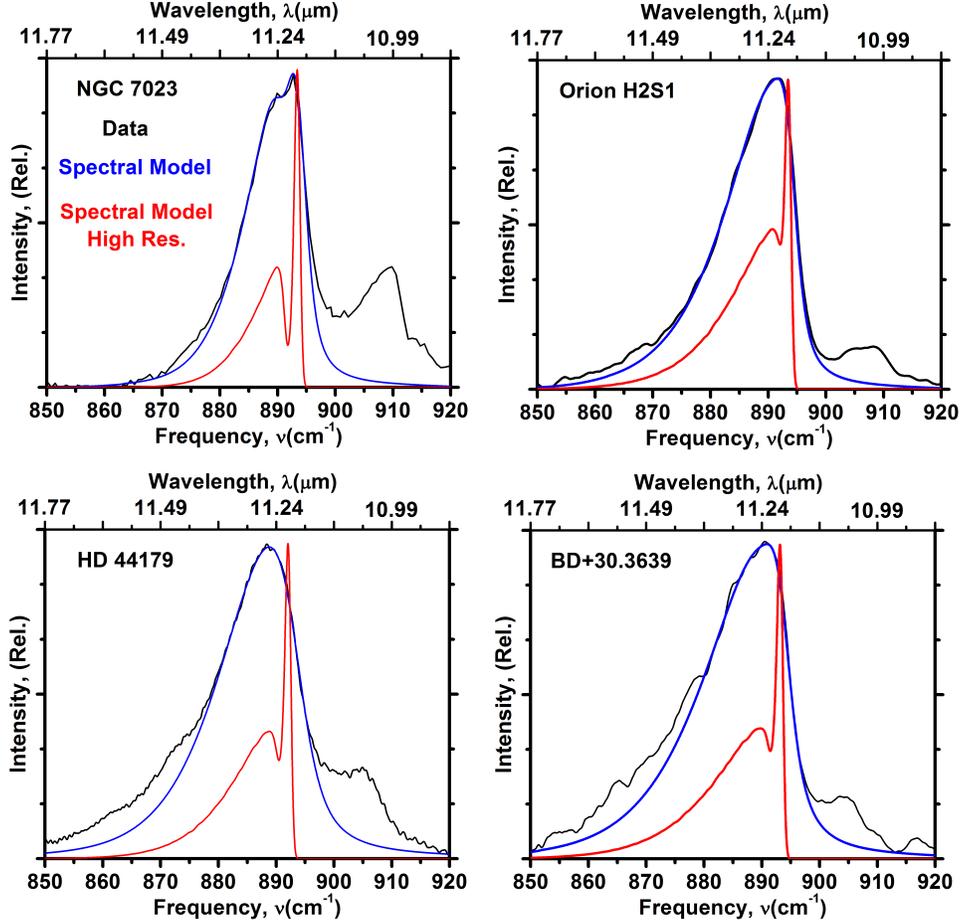

**Figure 5.** Spectral fits to the 11.2 μm UIR bands for different objects. In each case, the fit parameters for NGC 7027 were perturbed to fit the observed profile (see Table 1 for all the fit parameters). As discussed previously, we suggest that the well-defined feature at 905 cm$^{-1}$ and the weaker, overlapping features from ~880-850 cm$^{-1}$ are not associated with the primary band giving rise to the 11.2 μm UIR feature.

**Table 1** Molecular spectral fit parameters[a]

| Object | $\lambda_0$(μm) | $\nu_0$(cm$^{-1}$) | $T_{rot}$(K) | $B$(cm$^{-1}$)[b] | $\Delta B/B$(%) | $\gamma_G$(cm$^{-1}$)[c] | $\gamma_L$(cm$^{-1}$)[c] |
|---|---|---|---|---|---|---|---|
| NGC 7027 | 11.22 | 891.4 | 825 | 0.0155 | -0.363 | 2.6 | 3.7 |
| Orion H2S1 | 11.21 | 892.2 | 620 | 0.0155 | -0.484 | 2.6 | 2.6 |
| HD 44179 | 11.23 | 890.4 | 620 | 0.0155 | -0.403 | 5.0 | 5.0 |
| BD+30.3639 | 11.22 | 891.4 | 860 | 0.0155 | -0.379 | 2.8 | 4.1 |
| NGC 7023 | 11.22 | 891.4 | 420 | 0.0155 | -0.405 | 2.5 | 2.5 |
| NGC 7023 | 12.73 | 785.7 | 460 | 0.0155 | +0.081 | 2.7 | 5.7 |

**Notes.**
(a) The 11.2 μm band was modeled as an $A_{1g} \rightarrow E_{1u}$ transition for a spherical top with $D_{6d}$ symmetry
(b) $B$ is the rotational constant for the lower transition level
(c) FWHM

## 3. DISCUSSION

We consider the broader astrophysical implications of our identification of $C_{24}$ as the likely carrier of the 11.2 μm UIR band. This identification raises a number of questions that are addressed below, including: (1) is the SED from the many IR active modes for $C_{24}$ consistent with all the UIR features? (2) is $C_{24}$ a potentially useful interstellar probe for local physical conditions? (3) what challenges remain unanswered in this effort?, and (4) what future observational, modeling, and laboratory efforts can help test the plausibility of $C_{24}$ as the carrier of the 11.2 μm UIR band.

### *3.1 DFT Frequencies and IR Intensities*

In order to estimate a SED for $C_{24}$ we need to determine the frequencies and intensities of its IR active vibrational modes. We performed B3LYP/6-31G(d) DFT calculations for these quantities. The results are reported in Table 2. The optimized structure is a slightly oblate symmetric top with $D_{6d}$ symmetry and ground state rotational constants of B=0.0181 and C=0.0190 cm$^{-1}$. We used a DFT frequency scaling factor of 0.986 for frequencies below 2700 cm$^{-1}$, as determined by Russell et al. 2010, and we have also adopted their error estimate of the scaling factor, ±0.0334. The uncertainties in the DFT infrared intensities are not readily available. Based on comparisons by Bauschlicher et al. 1999 of B3LYP DFT calculations to laboratory IR absorption spectra, for several PAHs in the size range of interest here, we made crude estimates of the uncertainties of ~±25% and ~±50% for the stronger and weaker bands, respectively. The frequency and intensity error estimates are relevant to the next section where we compare DFT-based SED predictions to an entire UIR spectrum.

Our DFT frequencies and IR intensities are in reasonable agreement with those of Adjizian et al. 2016; however, one must account for the different frequency scaling factors used in the two studies. Adjizian et al. used a non-standard scaling factor chosen to match their $C_{60}$ calculation to an observed band location of $C_{60}$. Their scaling factor is 1.046, as compared to our value of 0.986.

From Table 2 we see that $C_{24}$ has 66 vibrational modes, of which 10 have significant infrared intensity (i.e., $A_{1,0}$>1 sec$^{-1}$). Four of these modes are doubly degenerate, resulting in six unique, strong IR bands for $C_{24}$. We show in the next section, that, to within the DFT uncertainties, all of the predicted bands correlate to observed UIR bands.

### *3.2 Modeled SED for $C_{24}$*

We estimated a SED for $C_{24}$ based on the method used by Pech et al. 2002 for PAH interstellar emission. Their approach approximates the instantaneous internal energy distribution of a radiating molecule in terms of an effective vibrational temperature, $T_{vib}$. The SED is constructed by summing computed $T_{vib}$-dependent spectra, each normalized to the same integrated energy, for equal energy decrements until the excitation energy is fully radiatively dissipated. The relationship between energy and $T_{vib}$ is determined using the standard formula for the vibrational heat capacity, and the result for $C_{24}$ is displayed in Figure 6. The electronically excited absorption spectrum of $C_{24}$ has not been measured, and we assume that the initial excitation is due to absorption of a 5 eV photon. The resultant SED is not overly sensitive to the initial excitation energy. A change of ±1 eV would change the effective vibrational emission temperature (i.e., ~$T_{vib}$ at the median energy) by ~± 100 K. The modeled SED is compared to the NGC 7027 UIR spectrum in Figure 7.

**Table 2**[a] DFT Band Origins and Einstein $A$'s[b] for $C_{24}$

| $\nu_0$(cm$^{-1}$) | $\lambda_0$(μm) | $A_{1,0}$(s$^{-1}$) | $\nu_0$(cm$^{-1}$) | $\lambda_0$(μm) | $A_{1,0}$(s$^{-1}$) |
|---|---|---|---|---|---|
| 272.4 | 36.71 | 0.00 | ***814.8*** | ***12.27*** | ***1.65*** |
| 274.6 | 36.42 | 0.00 | ***907.9*** | ***11.01*** | ***3.41*** |
| 399.0 | 25.07 | 0.00 | ***908.2*** | ***11.01*** | ***3.44*** |
| 403.4 | 24.79 | 0.12 | 994.7 | 10.05 | 0.00 |
| 404.1 | 24.75 | 0.12 | 995.9 | 10.04 | 0.00 |
| 472.2 | 21.18 | 0.00 | 1005.7 | 9.94 | 0.00 |
| 499.4 | 20.02 | 0.00 | 1006.2 | 9.94 | 0.00 |
| 499.7 | 20.01 | 0.00 | **1029.1** | **9.72** | **1.71** |
| 521.0 | 19.20 | 0.20 | 1043.9 | 9.58 | 0.00 |
| 521.2 | 19.19 | 0.21 | 1088.3 | 9.19 | 0.00 |
| 537.7 | 18.60 | 0.00 | 1093.3 | 9.15 | 0.00 |
| 540.6 | 18.50 | 0.00 | 1127.1 | 8.87 | 0.00 |
| **544.1** | **18.38** | **11.81** | 1127.7 | 8.87 | 0.00 |
| 575.0 | 17.39 | 0.00 | 1158.5 | 8.63 | 0.00 |
| 575.3 | 17.38 | 0.00 | 1172.1 | 8.53 | 0.05 |
| ***592.4*** | ***16.88*** | ***1.15*** | 1172.4 | 8.53 | 0.05 |
| ***593.3*** | ***16.85*** | ***1.17*** | 1189.0 | 8.41 | 0.00 |
| 615.0 | 16.26 | 0.00 | 1206.1 | 8.29 | 0.00 |
| 626.7 | 15.96 | 0.24 | 1251.5 | 7.99 | 0.05 |
| 626.9 | 15.95 | 0.00 | 1251.6 | 7.99 | 0.04 |
| 626.9 | 15.95 | 0.26 | ***1263.4*** | ***7.92*** | ***4.55*** |
| 627.1 | 15.95 | 0.00 | ***1263.5*** | ***7.91*** | ***4.59*** |
| 632.7 | 15.81 | 0.00 | 1320.8 | 7.57 | 0.00 |
| 721.3 | 13.86 | 0.00 | 1431.3 | 6.99 | 0.00 |
| 721.4 | 13.86 | 0.00 | 1431.5 | 6.99 | 0.00 |
| 733.8 | 13.63 | 0.00 | 1445.2 | 6.92 | 0.00 |
| 734.0 | 13.62 | 0.00 | 1445.5 | 6.92 | 0.00 |
| 735.4 | 13.60 | 0.02 | 1506.8 | 6.64 | 0.00 |
| 735.7 | 13.59 | 0.00 | 1507.3 | 6.63 | 0.00 |
| 736.9 | 13.57 | 0.00 | 1513.2 | 6.61 | 0.00 |
| 761.3 | 13.14 | 0.00 | 1520.1 | 6.58 | 0.00 |
| 785.3 | 12.73 | 0.10 | 1520.6 | 6.58 | 0.00 |
| **814.6** | **12.28** | **1.63** | 1541.6 | 6.49 | 0.0 |

**Notes.**

(a) Bands with $\underline{A_{1,0}}$>1 s$^{-1}$ highlighted in bold with nearly degenerate bands are italicized

(b) IR intensity unit conversion: $A(\text{sec}^{-1})=1.253 \times 10^{-7} \nu(\text{cm}^{-1})^2 S(\text{km/mol})$

The overall consistency between the modeled SED and the NGC 7027 UIR spectrum supports the idea that $C_{24}$ is a plausible carrier for the 11.2 μm feature and suggests that it is a major contributor to several other features. The 12.7 μm feature is considered in detail later, where, for NGC 7023, we find that $C_{24}$ is the dominant contributor to the observed UIR band. However, there are at least two additional, variable, overlapping contributions to this band, including a band feature at ~798 cm$^{-1}$ (12.5 μm) and a Ne II line at 12.8 μm. Peeters et al. 2002 have shown that the 7.7 μm UIR emission complex decomposes into two main components centered at 7.6 μm and 7.9 μm. The $C_{24}$ SED has a strong feature that could account for the 7.9 μm UIR component. UIR spectra in the 15-21 μm spectral region (Peeters et al. 2004) are complex and variable, consisting of several narrow bands perched on a broader plateau feature. The $C_{24}$ SED in this range also exhibits complex structure and could account for a significant fraction of the UIR emission.

### 3.2 Modeled SED for $C_{24}$

We estimated a SED for $C_{24}$ based on the method used by Pech et al. 2002 for PAH interstellar emission. Their approach approximates the instantaneous internal energy distribution of a radiating molecule in terms of an effective vibrational temperature, $T_{vib}$. The SED is constructed by summing computed $T_{vib}$-dependent spectra, each normalized to the same integrated energy, for equal energy decrements until the excitation energy is fully radiatively dissipated. The relationship between energy and $T_{vib}$ is determined using the standard formula for the vibrational heat capacity, and the result for $C_{24}$ is displayed in Figure 6. The electronically excited absorption spectrum of $C_{24}$ has not been measured, and we assume that the initial excitation is due to absorption of a 5 eV photon. The resultant SED is not overly sensitive to the initial excitation energy. A change of ±1 eV would change the effective vibrational emission temperature (i.e., ~$T_{vib}$ at the median energy) by ~± 100 K. The modeled SED is compared to the NGC 7027 UIR spectrum in Figure 7.

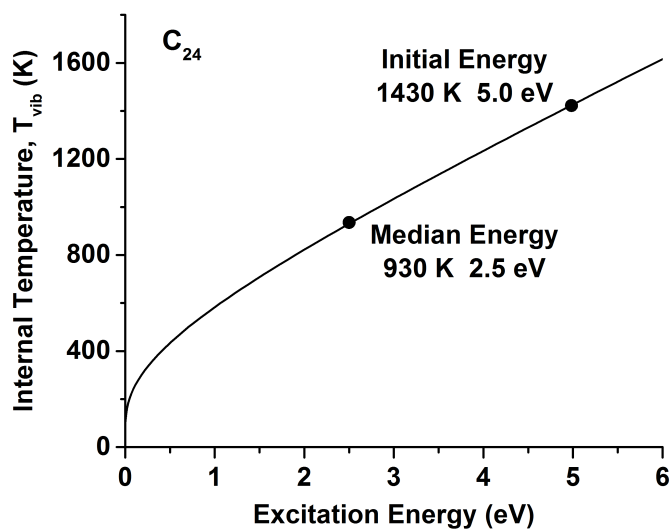

**Figure 6.** Effective vibrational temperature, $T_{vib}$, for $C_{24}$ as a function of its internal energy. This curve is based on the heat capacity for $C_{24}$ computed using the DFT vibrational energies. We assumed an excitation energy of 5 eV for estimating the SED, which corresponds to $T_{vib}$=1430 K.

The overall consistency between the modeled SED and the NGC 7027 UIR spectrum supports the idea that $C_{24}$ is a plausible carrier for the 11.2 μm feature and suggests that it is a major contributor to several other features. The 12.7 μm feature is considered in detail later, where, for NGC 7023, we find that $C_{24}$ is the dominant contributor to the observed UIR band. However, there are at least two additional, variable, overlapping contributions to this band, including a band feature at ~798 cm$^{-1}$ (12.5 μm) and a Ne II line at 12.8 μm. Peeters et al. 2002 have shown that the 7.7 μm UIR emission complex decomposes into two main components centered at 7.6 μm and 7.9 μm. The $C_{24}$ SED has a strong feature that could account for the 7.9 μm UIR component. UIR spectra in the 15-21 μm spectral region (Peeters et al. 2004) are complex and variable, consisting of several narrow bands perched on a broader plateau feature. The $C_{24}$ SED in this range also exhibits complex structure and could account for a significant fraction of the UIR emission.

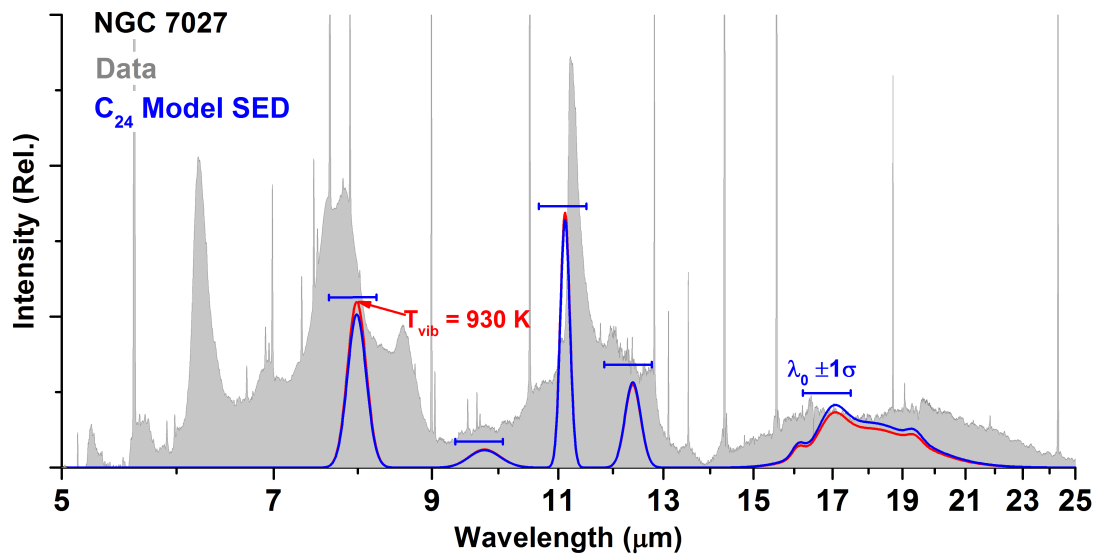

**Figure 7.** Comparison of the observed NGC 7027 UIR spectrum to the modeled SED for $C_{24}$, based on the DFT calculations (see Table 2). ). The model SED (blue curve) adopts Gaussian band shapes with the widths selected to approximate the observed widths for the corresponding spectral bands and areas proportional to the model intensities. The model widths (FWHM) are 0.28 (44), 0.57 (60), 0.19 (16), 0.36 (24), 0.36 (14), 0.80 (28), 3.38 (100), and 0.55 (14) μm (cm$^{-1}$) for the 7.91, 9.72, 11.01, 12.27, 15.95, 16.86, 18.38, and 19.91 μm bands, respectively. The horizontal bars centered on the modeled bands denote the estimated errors, ±0.035$\lambda_0$, in the DFT frequency calculations. The modeled bands were shifted to the red by $\Delta\lambda$=+0.01$\lambda$ to approximate the effect of anharmonic coupling at high $T_{vib}$. It is seen that the SED based on the model of Pech et al. 2002 is well approximated by the spectrum for the median energy, corresponding to $T_{vib}$=930 K (red curve).

*3.3 Spectral Modeling of the NGC 7027 12.7 μm UIR Band*

While we have focused on the 11.2 μm IR band, $C_{24}$ has many other active IR bands that should make significant contributions to the UIR spectrum. Our DFT calculations predict that the 12.7 μm UIR band should also originate from $C_{24}$. We have applied the same spectral analysis approach used for the 11.2 μm band to the 12.7 μm UIR band for NGC 7023. The results are displayed in Figure 8 and the fit parameter values are presented in Table 1. We selected NGC 7023 because it provided a much cleaner background-subtracted 12.7 μm band (Shannon et al. 2016) than could be obtained from NGC 7027. An interesting feature of the 12.7 μm band is that it displays a red band head and a blue tail, which means its $\Delta B$ must be positive. A positive $\Delta B$ is less common than a negative $\Delta B$ because most vibrational modes tend to slightly increase the vibrationally-averaged size of a molecule. But, polyatomic molecules have many modes ($C_{24}$ has 66), exhibiting a wide variety of symmetries, a few of which can have a positive $\Delta B$. We are currently collaborating with colleagues at the Leiden Observatory and NASA Ames (see Mackie et al. 2015) to perform more advanced DFT calculations to predict the $\Delta B$'s and anharmonic constants for the $C_{24}$ modes. These computations are more demanding than those for just the frequencies and IR intensities, discussed earlier. We will report the results when they become available. For now, we conclude that the spectral fit for the 12.7 μm band supports the idea that both the 11.2 and 12.7 μm UIR band originate from the same carrier, $C_{24}$.

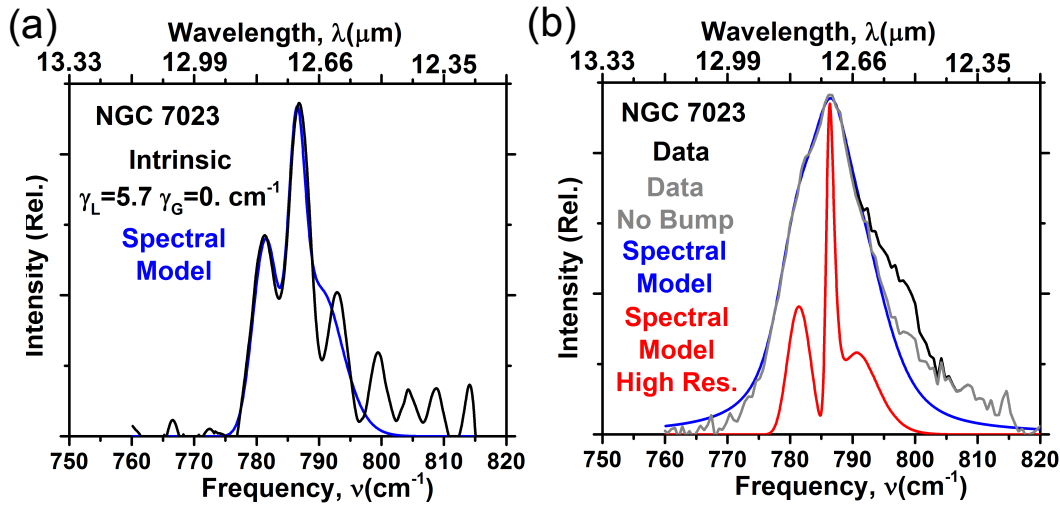

**Figure 8.** Spectral modeling of the NGC 7023 12.7 μm UIR band. **a.** Spectral fit to the intrinsic profile based on perturbing the fit parameters for the NGC 7023 11.2μm UIR band (see Table 1 for the fit parameters). The intrinsic profile was based on deconvolution of Lorentz broadening from the observed data with the feature at ~798 cm$^{-1}$ removed via subtraction of a Gaussian approximation of the feature (see Figure 8b). **b.** Comparison of spectral fit to original data (i.e., application of Lorentz broadening to the molecular spectral fit to the intrinsic profile). Also shown is a high spectral resolution of the modeled profile with most of the broadening removed ($\gamma_L$=0.0 and $\gamma_G$=1.0 cm$^{-1}$).

### 3.4 Challenges to the $C_{24}$ Carrier Identification

At this initial stage of investigation, $C_{24}$ is a viable carrier for the 11.2 μm UIR band. It provides an Occam's razor explanation for the band, requiring only a small number of spectroscopic and environment parameters needed to fit the band profile and its variations, and a SED consistent with the entire UIR spectrum. However, there are a number of challenges to address.

One concerns the established correlation, or lack thereof, between the 11.2 μm and other bands (Tielens 2008). For example, the relative equivalent widths of 11.2 and 3.3 μm bands are well correlated. This correlation is easier to explain with a PAH model because both bands are attributed to hydrogen bends and stretches on the same, neutral PAH molecules. On the other hand, $C_{24}$ does not contain any hydrogen, therefore the correlation must arise from a correlation of the abundances of $C_{24}$ and hydrocarbons, which exhibit a 3.3 μm band. Another challenge concerns the lack of correlation between the 11.2 and 12.7 μm bands. For the PAH model the lack of correlation is attributed to the argument that the 12.7 μm band arises from a variable mixture of neutral and charged PAHs (Shannon et al. 2016). On the other hand, $C_{24}$ exhibits strong 11.2 and 12.7 μm bands, implying their intensities should be strongly correlated. However, as discussed earlier, the 12.7 μm UIR band consists of three overlapping, independently variable components, with only one attributed to $C_{24}$. As invoked for PAHs, multiple, independently varying components should result in de-correlation of the observed 11.2 and 12.7 μm bands.

$C_{24}$ cannot account for the 6.2 μm UIR band because it has no IR active modes below 7.9 μm. Thus, a different molecular species is required. We are currently applying the deconvolution algorithm to the NGC 7027 6.2 μm band and will report our findings in a future paper. Briefly, our preliminary results show an intrinsic profile whose shape indicates a prolate top molecule of size and shape similar to Naphthalene ($C_{10}H_8$).

Another challenge concerns the large abundance, a few percent of the elemental carbon, implied by $C_{24}$ as the dominant carrier of the 11.2 μm band. Possible explanations include extreme stability in a harsh UV environment and/or efficient chemical formation. The photo-ionization energy for $C_{24}$ is predicted to be 7.7 eV (Tiago et al.2008). This is slightly larger than that measured for $C_{60}$, 7.6 eV, which is known to be an extremely stable interstellar molecule (Castellanos et al. 2014). In comparison, the photo-ionization energy for a large compact PAH, $N_C$=50, is about 6.1 eV (Montillaud et al. 2013). PAHs are also thought to be very stable interstellar species, thus, we anticipate that $C_{24}$, due to its significantly higher photo-ionization energy, has even greater stability. PAHs can also undergo photo-dissociation through loss of hydrogen, which can occur at a comparable rate to photo-ionization. In contrast, this photo-destruction pathway is not available to $C_{24}$.

A potential chemical formation pathway for $C_{24}$ is analogous to that already demonstrated for $C_{60}$ and other comparably large fullerenes (Zhen et al. 2014; Berne et al. 2015). The mechanism involves photo-induced modification of a large PAH (i.e., complete dehydrogenation and cleavage of one or more $C_2$'s) to form a bare carbon skeleton that folds up to form a fullerene.

The $C_{24}$ model does not provide an explanation for any of the broad plateau features that typically accompany the narrower UIR bands. Both the PAH and MAON (Mixed Aromatic-aliphatic Nanoparticles) (Kwok & Zhang 2011) models offer a plausible explanation for the plateaus in terms of a superposition of many broadened spectra. This explanation is not in conflict with the $C_{24}$ model. Our findings suggest a "grandPAH" type of explanation for the UIR bands in which a small number of robust and abundant molecules account for these features, and PAHs and MAONs can account for the plateau features.

### *3.5 Potential Utility of $C_{24}$ as an Interstellar Probe of Local Physical Conditions*

The variability of the $C_{24}$ fit parameters for different objects suggests that it may be a useful probe of local interstellar physical conditions. Inspection of the spectral fit parameters in Table 2 for the different objects reveals that $T_{rot}$ and the broadening widths, $\gamma_L$ and $\gamma_G$, exhibit the largest variations. $T_{rot}$ varies from 420 to 860 K. Based on our assumption that $T_{rot}=T_{H2}$ for NGC 7027, we would expect that the retrieved $T_{rot}$'s for the other objects would correlate strongly with their $T_{H2}$'s. Bernard-Salas and Tielens 2005 present $T_{H2}$'s for a wide variety of planetary nebulae, where $T_{H2}$ ranges from ~400-1000 K. Specific comparisons can be made for most of the objects in our study. For the Orion Bar, Shaw et al. 2009 found that $T_{H2}$ varies over the range of 400-1000 K. For NGC 7023, Fuente et al. 1999 found that $T_{H2}$ varies over the range of 300-700 K. And, for BD +30.3639, Shupe et al. 1998 determined that $T_{H2}$ varies over the range of 800-1000 K. These ranges reflect spatial variability in $T_{H2}$. The $T_{rot}$'s from our analysis fit within the observed ranges for $T_{H2}$ for these objects, warranting future studies to perform more spatially consistent comparisons.

The variability of $\gamma_L$ and $\gamma_G$ is attributed to the variations in the average $T_{vib}$ for each object. Both widths should increase with an increase in $T_{vib}$. The quantitative relationships of the widths with $T_{vib}$ for $C_{24}$ are not known. However, we can gain semi-quantitative insight into the relative variations of $\gamma_L$ between the different objects if we adopt the relationship established for compact PAHs (Candian & Sarre 2015),

$$\Delta\gamma_L = \chi'' \Delta T_{vib}, \qquad (4)$$

where $\chi''$=0.016 cm$^{-1}$/K and $\Delta T_{vib}$ is the difference in the average vibrational temperature between two objects. The largest difference in widths, $\Delta\gamma_L$=2.5 cm$^{-1}$, occurs between HD 44179 and NGC 7023, and implies a temperature difference of $\Delta T_{vib}$=156 K. Based on $C_{24}$ heat capacity calculations discussed earlier (see Figure 6), we can interpret this temperature difference in terms of a difference in the internal excitation energy of $\Delta E$=0.8 eV. This difference could arise from a difference in the energy of the dominant excitation process (i.e., absorption of a UV photon), from a difference in the internal energy of the ground electronic state, or a combination of both causes. The second possibility implies a $C_{24}$ ground electronic state vibrational temperature ~500 K hotter in HD 44179 than in NGC 7023. This large temperature difference seems plausible, in view of the optically opaque, high temperature (in excess of 1000 K at the inner dust radius), circumstellar dust layer in HD 44179 (Men'shchikov et al. 2002) that would radiatively heat the $C_{24}$. We could perform a similar analysis based on the established relationship between the observed band origin location, $\nu_0$, and $T_{vib}$; however, the results would be less meaningful because the spectral resolutions of the *Spitzer* IRS and ISO SWS sensors are comparable to the variations in $\nu_0$.

This discussion suggests that $C_{24}$ may be a useful probe for the local temperature and the local UV and IR radiation fields. We note that Brieva et al. 2016 discussed the use of the fullerene $C_{60}$ as a probe of astrophysical environments. They considered the relationship of IR band ratios with respect to local physical conditions. Our discussion above pertains to the information that can be obtained from the profile of the 11.2 μm band. As explored earlier, $C_{24}$ has other observable bands, which, by analogy to the $C_{60}$ band ratios, may also provide additional information on local physical conditions.

### 3.6 Recommendations for Future Efforts

Recommendations for future investigations to clarify the plausibility of $C_{24}$ as the carrier of the 11.2 μm UIR band include:

1. DFT calculations for the $C_{24}$ anharmonic vibration and vibration-rotation constants: Recently, Mackie et al. 2015 demonstrated remarkable agreement between laboratory IR spectra and DFT calculations, including anharmonicity, for several PAHs. This agreement was obtained without the use of scaling factors to approximately correct DFT harmonic frequencies to the observed frequencies. Application of the anharmonic DFT approach to $C_{24}$ would narrow the uncertainties of our SED prediction, resulting in a more definitive evaluation of the plausibility of $C_{24}$ as a UIR carrier. The DFT calculations could also confirm (or disprove) our prediction of an unusual, positive vibration-rotation constant for the 12.7 μm band.

2. Laboratory synthesis and IR spectral measurements of $C_{24}$: These measurements would provide a definitive determination of the plausibility of the $C_{24}$ carrier hypothesis.

3. Deconvolution of other UIR and DIB bands: Irrespective of the validity of the $C_{24}$ carrier hypothesis, our deconvolution approach can be used to derive intrinsic profiles for all the strong and weak UIR bands, and thereby provide tighter constraints on proposed carriers.

We anticipate that it will also aid in the identification of DIB carriers, as these bands also exhibit significant broadening effects.

4. Observational testing of the $T_{rot}$-$T_{H2}$ correlation: We predict a tight correlation between $T_{rot}$ for $C_{24}$ and $T_{H2}$. Our limited consideration of this correlation is encouraging, but it is far from definitive, due to lack of spatially consistent comparisons, within an object, between the two temperatures. Relevant archival data may exist, and we would greatly appreciate being informed of such data. It may be possible to perform spatially correlated measurements using ground-based or airborne (i.e., SOFIA) observatories. The enhanced spatial resolution of the James Web Space Telescope (launch date October 2018) will facilitate these measurements.

## 4. SUMMARY AND CONCLUSIONS

We have deconvolved the observed spectrum of the 11.2 μm UIR emission band for NGC 7027. The resulting intrinsic profile strongly indicates that the carrier molecule is small ($N_C < 30$), and nearly spherical. Further refinement of the analysis reveals that the fullerene $C_{24}$ is the likely molecule. This identification is supported by DFT modeling of $C_{24}$. Our approach is applied to the 11.2 μm band in several other sources, further bolstering the case for $C_{24}$.

The molecular carrier was identified as a small fullerene, $C_{24}$, based on the key assumption that the UIR profile is broadened due to a combination of Lorentz/lifetime and Gaussian/anharmonicity broadening. A spectral deconvolution algorithm was used to remove the effects of broadening. This revealed an intrinsic spectral profile with a distinct shape characteristic of a molecular vibration-rotation profile with as sharp R-branch feature and a broader red shaded tail. Applying a well-established spectroscopic idea, that the fractional area of the R-branch is a direct measure of molecular geometry, we determined that the carrier had to be close to spherical. Because of its high molecular symmetry (i.e., no dipole moment), the carrier's rotational temperature should be equilibrated with the gas kinetic temperature, as measured by $T_{H2}$ (molecular hydrogen rotational temperature). Based on the $BT_{rot}$-UIR band width constraint (eq.(3)), with $T_{rot}=T_{H2}=825$ K for NGC 7027, we derived an upper limit size to the carrier of $N_C<30$. Given the above size and shape constraints, the only possible carrier candidate is the fullerene $C_{24}$. A spectral fit to the NGC 7027 11.2 μm band intrinsic profile corroborated the carrier identification. We performed a DFT calculation to estimate the frequencies and IR intensities of the $C_{24}$ vibrational modes. The DFT results were used to predict a SED for $C_{24}$ that was found to be consistent with the entire NGC 7027 UIR spectrum.

We conclude that $C_{24}$ is a plausible carrier for the 11.2 μm UIR band as well as for part or all of several other UIR bands. We speculate that it may be an abundant interstellar molecule due to its high photo-ionization energy. Finally, if $C_{24}$ is a UIR band carrier, then it will become a useful probe of astrophysical environments.

L.B. an R. S. appreciate funding from Spectral Sciences, Inc., under an internal research and development project. The authors thank M. Shannon, E. Peeters (University of Western Ontario), and S. Kwok (University of Hong Kong) for sharing UIR data files, J. Cami (University of Western Ontario) for interactions regarding PAHs, and J. Gelbord (Spectral Sciences, Inc.) for review of the draft manuscript. We thank the anonymous reviewer for insightful comments and suggestions for improving the paper.